\documentclass[fleqn,usenatbib]{mnras}

\usepackage{newtxtext,newtxmath}
\usepackage[T1]{fontenc}
\usepackage{ae,aecompl}
\usepackage[normalem]{ulem}

\usepackage{graphicx}	% Including figure files
\usepackage{amsmath}	% Advanced maths commands
\usepackage{amssymb}	% Extra maths symbols

 \graphicspath{{./../files/}}

\title[HD 141569]{ALMA and VLA Observations of the HD 141569 System}
\author[White et al.]{Jacob Aaron White$^{1}$\thanks{E-mail: jawhite@astro.ubc.ca},
A.C. Boley,$^{1}$
M.A. MacGregor,$^{2, 3}$
A.M. Hughes,$^{4}$ D.J.
\newauthor
Wilner$^{2}$
\\
$^{1}$Department of Physics and Astronomy, 6224 Agricultural Road, Vancouver, BC V6T 1Z1, CAN\\
$^{2}$Harvard-Smithsonian Center for Astrophysics, 60 Garden Street, Cambridge, MA 02138, USA\\
$^{3}$Department of Terrestrial Magnetism, Carnegie Institution for Science, 5241 Broad Branch Road, Washington, DC 20015, USA\\
$^{4}$Department of Astronomy, Van Vleck Observatory, Wesleyan University, 96 Foss Hill Dr., Middletown, CT 06459, USA
}

\date{Accepted XXX. Received YYY; in original form ZZZ}

\pubyear{2016}

\begin{document}
\label{firstpage}
\pagerange{\pageref{firstpage}--\pageref{lastpage}}
\maketitle

% Abstract of the paper
\begin{abstract}

We present VLA 9 mm (33 GHz) observations of the HD 141569 system from semester 16A. The observations achieve a resolution of 0.25 arcsec ($\sim28$ au) and a sensitivity of $4.7~\mu \rm Jy~beam^{-1}$. We find (1) a $52\pm 5~\mu$Jy point source at the location of HD 141569A that shows potential variability, (2) the detected flux is contained within the SED-inferred central clearing of the disc meaning the spectral index of the dust disc is steeper than previously inferred, and (3) the M dwarf companions are also detected and variable. Previous lower-resolution VLA observations (semester 14A) found a higher flux density, interpreted as solely dust emission.  When combined with ALMA observations, the VLA 14A observations suggested the spectral index and grain size distribution of HD 141569's disc was shallow and an outlier among debris systems. Using archival ALMA observations of HD 141569 at 0.87 mm and 2.9 mm we find a dust spectral index of $\alpha_{\rm mm} = 1.81\pm 0.20$. The VLA 16A flux corresponds to a brightness temperature of $\sim5\times10^{6}$ K, suggesting strong non-disc emission is affecting the inferred grain properties. The VLA 16A flux density of the M2V companion HD 141569B is $149\pm9~\mu$Jy, corresponding to a brightness temperature of $\sim2\times10^{8}$ K and suggesting significant stellar variability when compared to the VLA14A observations, which are smaller by a factor of $\sim6$. 
\end{abstract}

% Select between one and six entries from the list of approved keywords.
% Don't make up new ones.

%%%%%%%%%%%%%%%%%%%%%%%%%%%%%%%%%%%%%%%%%%%%%%%%%%

%%%%%%%%%%%%%%%%% BODY OF PAPER %%%%%%%%%%%%%%%%%%

\section{Introduction}

HD 141569 is a multiple star system that contains an extensive circumstellar disc of dust and gas.  The disc was first detected by \textit{IRAS} \citep{walker,andrillat} and later imaged in scattered light by  {\it HST}. The scattered light images revealed that the disc has large spiral/ring structures at separations of $\sim 100$ to 400 au \citep{ancker,fisher,weinberger, clampin}.  Simulations by \cite{augereau_papaloizou} showed  that some of these morphological features are consistent with perturbations by two nearby ($\sim7.5$ arcsec) M dwarfs \citep{weinberger,reche}, but suggested that planetary perturbations may also be required to explain the morphology.
 
Interior to the spiral structure is an additional dust disc, which  has been partly imaged in scattered light \citep{konishi} and resolved by ALMA at 870 $\mu$m (345 GHz) \citep{white}.  We refer to this component as the ``warm'' disc and the outer rings/spirals as the ``cold'' disc to distinguish between these morphologically separated areas.  The scattered light and ALMA observations show that the outer radius of the warm disc is at about 55 au, leaving a large gap between the warm and cold components.  This further suggests that an embedded object, such as a planet or planets, may be responsible for some of the disc structure.  Furthermore, SED modelling predicts that the warm disc has an inner radius of 11 to 17 au  \citep{malfait, maaskant}, which again, may require the presence of an unseen companion \citep{brittain}.  
 
There is also a resolved gas disc, observed in CO emission, that extends from about 30 to 210 au  \citep{flaherty,white}, straddling the warm and cold components. The origin of this gas is debated in the literature,  i.e., whether it is leftover from the disc's formation or being released from material within the disc \citep[such as the dynamical evolution of comets; e.g.,][]{dent}.  For a more extensive overview of the previous gas and debris observations of HD 141569, see \citet{white}.  

In addition to the large-scale morphological features, HD 141569's disc is of interest because of its potentially unique grain size distribution.  Previous \textit{Karl G. Jansky Very Large Array} (VLA) 9 mm (33 GHz) observations from Semester 14A \citep{macgregor}  measured an unresolved flux density of $85 \pm 5~\mu$Jy centred on HD 141569A and its disc (the M dwarf subsystem was resolved separately).  This emission was attributed mainly to the warm disc due to the expectation that the flux density from HD 141569A would be negligible.  When combined with shorter wavelength observations, the 9 mm flux density implies that the warm disc must have a spectral slope $\alpha_{mm} = 1.63\pm 0.06$.  If the emission is indeed due to dust in the warm disc, the result implies that the corresponding grain size distribution is very shallow and a clear outlier compared with debris discs \citep{macgregor}. If this interpretation is correct, then HD 141569's disc may reflect a unique (or short lived) evolutionary state.  There may however be other reasons for the large 9 mm flux density.  The beam size for the semester 14A observations was $3.0\times2.4$ arcsec$^2$  ($\sim300$ au at the system distance), which may include diffuse emission from the cold disc that was missed in some of the higher frequency, higher resolution observations.  However, this would still be difficult to reconcile with typical spectral slopes.  

Another possible source of contamination is HD 141569A itself. HD 141569A is  a Herbig Ae/Be pre-main sequence star classified as   B9.5Ve by \cite{jaschek} and A2Ve, as well as a possible $\lambda$ Boo member, by \citet{murphy2015}. The $\lambda$ Boo features suggest that HD 141569A is accreting dust-poor gas from its disc \citep{venn}.  \citet{mendigutia} detect H$_{\alpha}$ emission coming from within $\sim0.12$ au of HD 141569A, but were unable to constrain whether or not the emission is from accretion. The atmospheres around Herbig Ae/Be stars are largely unstudied observationally at millimetre/centimetre (mm/cm) wavelengths and the corresponding emission may not be captured well in stellar models.  Moreover, the submm and cm emission from main sequence stars other than the Sun are just now being explored, e.g. $\alpha$ Cen \citep{liseau}; AU Mic \citep{macgregor13}; Fomalhaut \citep{white_fom}; Vega \citep{hughes}; Sirius A \citep{white_sir}.  If significant non-disc emission is present in pre-MS and MS stars, then inferred dust properties in circumstellar discs may be biased by this emission. 

The possibility of significant mm/cm emission from stars is highlighted by HD 141569A's  M dwarf companions, HD 141569B and HD 141569C.  The stars were not detected by ALMA at 870 $\mu$m (345 GHz) \citep{white} or 2.9 mm (103 GHz) (as discussed here), but \citet{macgregor} found a 9 mm flux of $51\pm5~\mu$Jy for both stars (the beam size was larger than the separation between the two stars). This flux is significantly greater than what would be expected from a blackbody at the photosphere temperature and is indicative of significant coronal processes that dominate the mm/cm wavelength emission for M dwarfs, as appears to be occurring in AU Mic \cite[e.g.,][]{macgregor13}.  

In order to properly characterize circumstellar debris, the emission from the host star must be taken into account. The observations presented here contribute to the ongoing project entitled \textit{Measuring the Emission of Stellar Atmospheres at Submillimetre/millimetre wavelengths} (MESAS). The MESAS project aims to assess the contributions of stellar atmospheres at submm-cm wavelengths and use the results to inform stellar atmosphere models \cite[e.g., PHOENIX][]{hauschildt}.

In this paper we present 9 mm observations of the HD 141569 triple system. Throughout this work, we adopt the \textit{Gaia} parallactic distance of $111\pm5$ pc \citep{gaia}. In Section 2 we give an overview of the observations. This includes the new VLA observations presented here, the archival VLA (semester 14A) observations, and also ALMA 2.9 mm measurements taken from the ALMA archive. In Section 3 we describe the model fitting procedures used to derive flux values. In Section 4 we discuss properties of the warm disc. In Section 5 we discuss the stellar atmospheres, and summarize the results in Section 6.

\section{Observations}

Observations were taken during the VLA Semester 16A (project ID VLA/2015-06-140) using the Ka band (9 mm) in the B antenna configuration with 27 antennas.  The longest baseline was 11.1 km. The observations were centred on HD 141569A using J2000 coordinates RA = 15 h 49 min 57.73 s and $\delta = -03^{\circ} 55' 16.62''$. Three scheduling blocks (SBs) were requested starting on 2016 June 19$^{\rm th}$, but only one SB was obtained, yielding a total on-source integration time of 1.15 hr. 

The correlator setup used $4\times2048$ MHz basebands with rest frequency centres at 30, 32, 34, and 36 GHz. Quasars J1256-0547 (3C279), J1246-0730, and J1557-0001 were used for bandpass and gain calibration. J1331+305 (3C286) was used as a flux calibration source. Data were reduced using the Common Astronomy Software Applications ({\scriptsize CASA 4.5.0}) pipeline \citep{casa_reference}, which included  bandpass, flux, and phase calibrations. The size of the synthesized beam is $0.29 \times 0.21$ arcsec$^{2}$ at a position angle of 21.6$^{\circ}$. The beam size corresponds to $\sim 28$ au at the system distance of 111 pc.  

\begin{figure}
\centering
\includegraphics[width=0.5\textwidth]{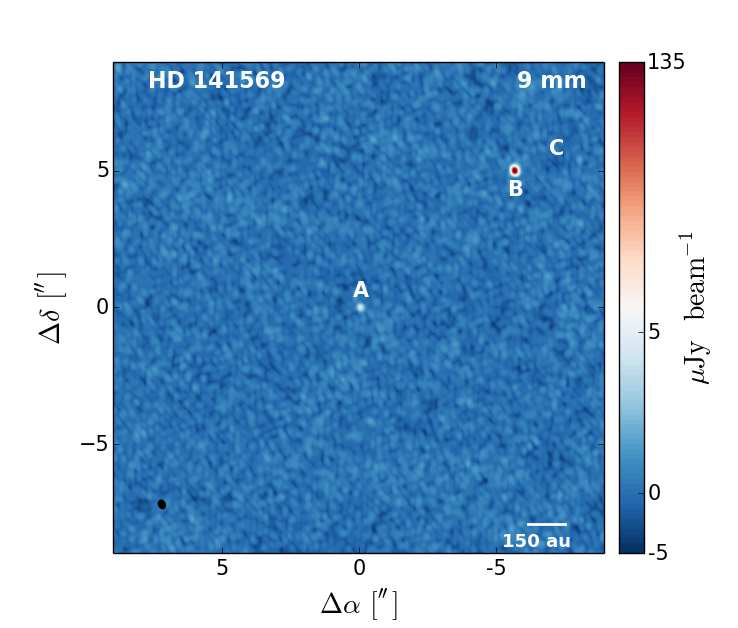}
\caption{CLEANed VLA 16A data of the HD 141569 system. The emission from HD 141569A and HD 141569B are marked by the letters ``A" and ``B". The expected \textit{Gaia} location for HD 141569C is marked by ``C". The synthesized beam is given by the black ellipse in the bottom left of the image and a 150 au ($\sim1.35$ arcsec) scale is given in the bottom right. Coordinates are given as offset from the phase centre. North is up and East is to the left.
\label{fig1}}
\end{figure}

Fig.\,\ref{fig1} shows the resulting 9 mm continuum image using natural weighting and cleaned using {\scriptsize CASA}'s \textit{CLEAN} algorithm  down to a threshold of $\frac{1}{2}~ \sigma_{\rm rms}$. The observations achieve a 0.25 arcsec resolution and a sensitivity of $4.7~\mu \rm Jy~beam^{-1}$. The average wavelength across the frequency range is 9.06 mm. HD 141569A and the companion HD 141569B were clearly detected. However, the disc around HD 141569A, as well the second companion, were not detected. The positions of all three stars are based on \textit{Gaia} astrometry and are denoted by A, B, and C in Fig.\,\ref{fig1}.

\subsection{Archival Data}
 
In addition to the new VLA 16A 9 mm observations presented here (henceforth referred to as VLA 16A), three other data sets are used in this study.  This includes VLA 9 mm observations from semester 14A (henceforth VLA 14A), which measured an unresolved flux density of $85\pm5 ~\mu$Jy centred on HD 141569A and its disc  \citep{macgregor}. As described below, we take this data from the VLA archive so that we can perform additional analysis of the observations. We take ALMA and SMA 870 $\mu$m flux and beam size values from literature \cite{white, flaherty}. We also present archival ALMA 2.9 mm observations that to our knowledge are not yet published. All of the newly present data and values from literature are summarized in Table \ref{obs}.
 
The VLA 14A data were taken from the VLA archive and calibrated with {\scriptsize CASA 4.5.0} pipeline \citep{casa_reference} in the procedure described in \citet{macgregor}. The observations took place on 2014 June 6$^{\rm th}$ in the D configuration with 25 antennas and a longest baseline of 1.31 km. The on-source integration time was 1.03 hr. Quasars J1256-0547, J1557-0001, and 3C286 were used as bandpass, gain and flux calibrators. The size of the synthesized beam is $3.0 \times 2.4$ arcsec$^{2}$ at a position angle of 338.6$^{\circ}$. The beam size corresponds to $\sim 300$ au at the system distance of 111 pc. These data achieve a sensitivity of $4.9~\mu \rm Jy~beam^{-1}$. 

The 2.9 mm ALMA data were taken from the ALMA archive (ID 2013.1.00883). These observations were made in two execution blocks (EBs) on 2015 August 8$^{\rm th}$, but one EB was not used due to phase amplitude and water vapor radiometer (WVR) problems. The total integration time for the successful EB was 1.24 hr.  Data were taken with 44 antennas with baselines ranging from 15 to 1547 m. Three different spectral windows were used with 2000 MHz bandpasses at rest frequencies centred at 98.24, 100.04, and 108.14 GHz. The correlator in FDM yielded 128 channels with widths of 15625 kHz. Data were reduced using {\scriptsize CASA 4.5.0} \citep{casa_reference} pipeline, which included WVR calibration; system temperature corrections; and bandpass, flux, and phase calibrations with quasars J1517-2422, J1550+054, and J1550+0527. Ceres was originally selected as a flux calibrator but the ALMA Pipeline quality assurance (QA) suggested to use the quasar J1550+05, as it is a more reliable calibrator for Band 3 wavelengths. 

\begin{figure}
\centering
\includegraphics[width=0.5\textwidth]{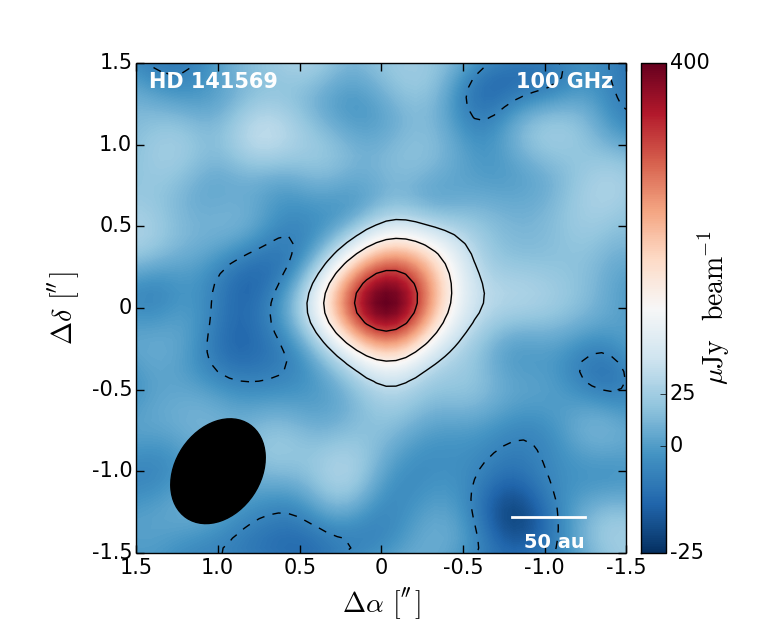}
\caption{CLEANed ALMA 2.9 mm data of the HD 141569 system. Solid contours show 3, 6, and $12\times \sigma_{\rm rms}$ and the dashed contour is $-\sigma_{\rm rms}$. The synthesized beam is given by the black ellipse in the bottom left of the image and a 50 au ($\sim0.45$ arcsec) scale is given in the bottom right. Coordinates are given as offset from the phase centre. North is up and East is to the left.
\label{fig_alma}}
\end{figure}

Figure \ref{fig_alma} shows the ALMA 2.9 mm continuum image of HD 141569. The companions HD 141569B and HD 141569C were not detected. The image was generated using natural weighting and cleaned using {\scriptsize CASA}'s \textit{CLEAN} algorithm  down to a threshold of $\frac{1}{2}~ \sigma_{\rm rms}$. The observations achieve a sensitivity of $24~\mu \rm Jy~beam^{-1}$. The size of the resulting synthesized beam  is $0.69 \times 0.52$ arcsec$^{2}$ at a position angle of 56$^{\circ}$, corresponding to $\sim 67$ au at the system distance of 111 pc.

\section{UV Modelling}

\subsection{VLA Data}

We use the {\scriptsize CASA} task \textit{uvmodelfit}  to recover the best fit flux density and position of HD 141569A and HD 141569B. The task fits a given component type to the visibilities of the data, which in this case is a single point source. The algorithm converges on the minimum $\chi^{2}$  iteratively.  The solution is sensitive to the starting position and flux values, which were taken from the CLEANed images. The sensitivity to the initial starting position allows us to fit point source models to components A and B separately.  The results are consistent with the peak flux densities per beam as measured directly from the dirty image, which for true point sources, should be the same value for the flux densities. Table \ref{fit_par} summarizes the best fit values for HD 141569A. We find a flux of $82 \pm 6~ \rm\mu Jy$ and $53\pm 5~\rm \mu Jy$ for VLA 14A and VLA 16A, respectively. The uncertainties represent the $\sigma_{\rm rms}$ and a 5\% flux calibration uncertainty, added in quadrature. The best fit value of $82 \pm 6~ \rm\mu Jy$ for VLA 14A agrees well with the $85 \pm 5~ \rm\mu Jy$ found by \citet{macgregor}. In the analysis, we use the value derived here.

HD 141569B was also fit with a point source model in the VLA 16A observations. The best fit flux is $149 \pm 9~ \rm\mu Jy$ and model results are summarized in Table \ref{fit_par}. The uncertainties represent the $\sigma_{\rm rms}$ and a 5\% flux calibration uncertainty\footnote{The 5\% flux calibration uncertainty was added in accordance with the VLA documentation detailed on https://science.nrao.edu/. If the absolute flux uncertainty is assumed to be negligible, then the uncertainties are characterized by $\sigma_{\rm rms}$ which is $4.9~\mu \rm Jy~beam^{-1}$ for VLA 14A and $4.9~\mu \rm Jy~beam^{-1}$ for VLA 16A.}, added in quadrature. As the separation between HD 141569B and HD 141569C is not resolved in the VLA 14A observations, we use $51 \pm 5~ \rm\mu Jy$ reported by \citet{macgregor} and assume the flux is equally distributed over the two companions.

\begin{table*}
\caption{Summary of {\scriptsize CASA}'s \textit{uvmodelfit} results. The algorithm converges on the minimum $\chi^{2}$  iteratively.  The solution is sensitive to the starting position and flux values, which were taken from the CLEANed images. The sensitivity to the initial starting position allows us to fit point source models to components A and B separately. A point source model was used for HD 141569A and HD 141569B in the VLA data. A disc model was used for the ALMA data. The uncertainties given by \textit{uvmodelfit} are not used throughout the analysis as they can be underestimated up to a factor of $\sqrt{\chi^{2}_{\rm reduced}}$. The location of the model is given as an offset from the phase centre of the observations. The uncertainty in the location is the statistical uncertainty in the model fitting procedure.} 
\centering 
\begin{tabular}{c  c  c  c || c } 
\hline\hline 

     & \multicolumn{3}{c||}{HD 141569A} & HD 141569B \\
   \hline
   	Parameter & VLA 16A 9 mm & VLA 14A 9 mm & ALMA 2.9 mm & VLA 16A 9 mm\\
   	\hline 
     %Starting Values [$\rm \mu Jy$, arcsec, arcsec] & [50, 0, 0] & [150, -5.7, 5] \\
     Flux [$\rm\mu Jy$] & $53\pm 3$ & $82\pm3$ & $419\pm 18$ & $149\pm 9$\\
     Location [mas] & $-21\pm5$, $-22\pm6$  & $91\pm37$, $-51\pm47$ &$13\pm19$, $12\pm10$ & $5670\pm2$, $4975\pm2$\\
     Major Axis [arcsec] & - & - & $0.64\pm 0.10$ & -\\
     Axis Ratio & - & - & $1.00\pm 0.21$ & -\\
     Position Angle [$^{\circ}$] & - & - & $19.7\pm57.3$ & -\\
     Reduced $\chi^{2}$ & 3.88616 & 5.67236 & 1.47999 & 3.88616\

\label{fit_par}
\end{tabular}
\end{table*}

\subsection{ALMA Data}

The archival ALMA 2.9 mm observations have a synthesized beam that is only slightly smaller than the warm disc.  For this reason, we use a disc with a flat intensity profile (instead of point source) to model the visibilties with the  {\scriptsize CASA} task \textit{uvmodelfit}.  The best fit flux density is $420~\mu$Jy. Including an absolute flux uncertainty of 10\% and the $\sigma_{\rm rms} = 24 ~\mu \rm Jy~beam^{-1}$ we find a total flux of $420\pm 50~\mu$Jy. The \textit{uvmodelfit} best fit parameters are also summarized in Table \ref{fit_par}.

\section{The Warm Dust Disc}

\subsection{The VLA 16A Null Detection of the Disc}

The warm disc is predicted by SED modelling to have an inner radius somewhere between 11-17 au \citep{malfait, maaskant} at the \textit{Gaia} distance of $111\pm5$ pc \citep{gaia}. The synthesized beams of the ALMA observations at 870 $\mu$m and 2.9 mm were too large to spatially resolve this central cavity. The synthesized beam for the VLA 14A observations was larger than the full extent of the warm disc, and thus also could not resolve the central clearing.   

The VLA 16A observations presented here have a synthesized beam width of 0.25 arcsec, which is sufficient for spatially resolving the warm disc's inner edge for radii $>14$ au.  The 870 $\mu$m ALMA observations found a peak intensity of 1740 $\mu \rm Jy~beam^{-1}$, which is presumed to be solely due to dust emission in the warm disc.  If this intensity is scaled to the frequency and beam size of the 9 mm observations using the spectral index of $\alpha_{mm}$ = 1.63 \citep{macgregor}, then the disc should have had a peak of $\sim22 ~ \mu \rm Jy~beam^{-1}$.

\begin{figure}
\centering
\includegraphics[width=0.5\textwidth]{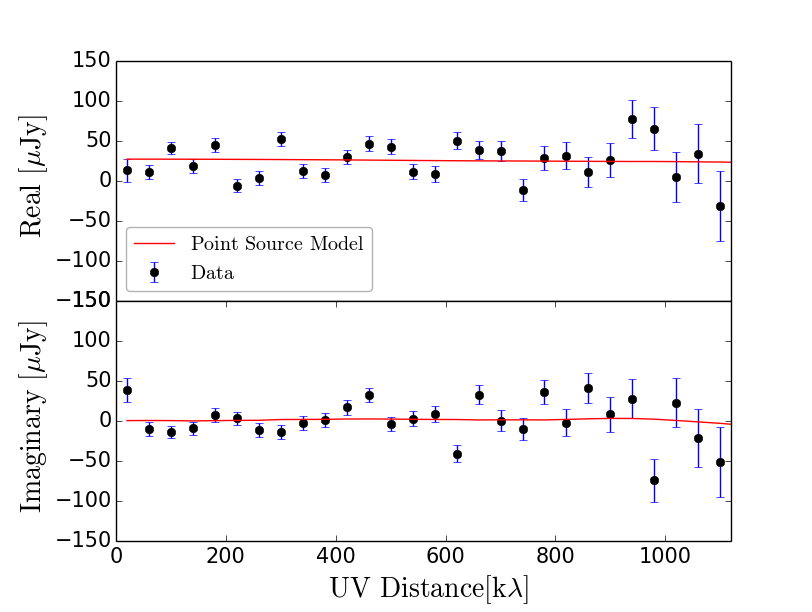}
\caption{Visibility plot of HD 141569A from VLA 16A. The top panel shows the real component of the visibilities and the bottom planel shows the imaginary component. The red line is the best fit point source model from \textit{uvmodelfit}. The data was annularly averaged with 40-k$\lambda$ bins. The uncertainties shown are the standard deviation of each 40-k$\lambda$ bin. The observations are consistent with a point source centred on the location of HD 141569A.
\label{vis_plot}}
\end{figure}

Fig.\,\ref{vis_plot} shows the visibilities of the VLA 16A observations. The projected visibilities shown are annularly averaged with 40-k$\lambda$ bins. Both the visibilities of the 16A observations and the reconstructed image are consistent with a point source centred within the SED predicted central clearing of the warm disc. With no detection of a warm dust disc, the observations place a $3\sigma$ upper limit on the intensity of the disc's inner edge to be $\lesssim 15~ \mu \rm Jy~beam^{-1}$ at 9 mm.

\subsection{Millimetre Spectral Index of the Disc}

\begin{table*}
\caption{Summary of best fit flux, peak emission, and beam size at each wavelength used in the analysis of the emission centred around HD 141569A. References are given for data taken from literature. The best fit flux from VLA 14A presented here is consistent with the $85 \pm 5~ \rm\mu Jy$ reported by \citet{macgregor}.}
\centering % used for centering table 
\begin{tabular}{c | c | c | c | c | c | c | c } % centered columns (3 columns) 
\hline\hline %inserts double horizontal lines 

   	Observatory & $\lambda$ [mm] & Flux [$\mu$Jy]& Peak [Jy beam$^{-1}$]& Beam Size [arcsec$^{2}$] & Beam P.A & Resolved & Reference \\
   	\hline 
	SMA & 0.87 & $8200\pm2400$ & $4100$ & $1.66\times1.16$ & $-82.2^{\circ}$ & N & \citet{flaherty} \\
	ALMA & 0.87 & $3800\pm500$ & $1740$ & $0.42\times0.34$ & $-61.1^{\circ}$ & Y & \citet{white} \\ 
	ALMA & 2.9 & $420\pm50$ & $350$ & $0.69\times0.52$ &$56.0^{\circ}$ & Y & - \\
	VLA 14A & 9  &  $82 \pm 6$ & $75$ & $3.0\times2.4$ &$338.6^{\circ}$ & N & -\\
	VLA 16A & 9 &  $53 \pm 5$ & $43$ & $0.29\times0.21$ & $21.6^{\circ}$ & N & - \\

\label{obs}
\end{tabular}
\end{table*}

\begin{figure}
\centering
\includegraphics[width=0.5\textwidth]{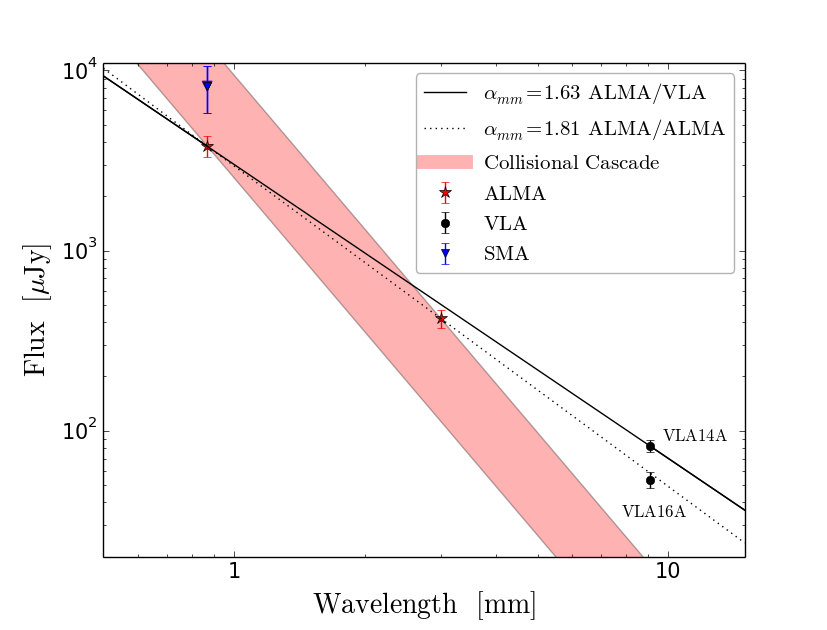}
\caption{Spectrum of HD 141569, showing the range of spectral indices based on different values for the flux densities. The ALMA observations are denoted by red stars. The VLA 9 mm observations from semesters 14A and 16A are denoted as black circles. The SMA 870 $\mu$m flux observation is denoted by the blue triangle. The red shaded area represents the expected flux values for a collisional cascade model (q = 3.5) with the flux anchored at the ALMA 870 $\mu$m and 2.9 mm observations. The calculations for $\alpha_{mm}$ are given in Section 4.2.  The uncertainties are for the total uncertainty of each observation ($\sigma_{\rm rms}$ and $\sigma_{\rm abs}$ in quadrature).
\label{spec}}
\end{figure}

Fig.\,\ref{spec} shows the flux densities as measured by ALMA (red), VLA (black), and SMA (blue).  Using the ALMA 870 $\mu$m and  VLA 14A 9 mm observations, \citet{macgregor} found a mm spectral index of $\alpha_{mm}=1.63$ (black line), assuming the flux density at these frequencies is characterized by  $F_\nu \propto \nu^{\alpha_{mm}}$.  Constraining the spectral index is key as it is related to the distribution of grain sizes over the given wavelength range \citep[e.g.,][]{wyatt}.  Adopting the methods of \citet{dalessio}, \citet{ricci}, \citet{macgregor}, and \citet{white_fom}, we can determine the slope of the grain size distribution, $q$, as follows. If the number of grains per size interval is given by $dn/dD \propto D^{-q}$ for grain diameter $D$, then the slope of the size distribution is related to the flux density spectral index by
\begin{equation}
q = \frac{\alpha_{mm} - \alpha_{pl}}{\beta_{s}} + 3,
\end{equation}
where $\beta_{s} = 1.8\pm0.2$ is a power law index for the dust opacity \citep{draine} and $\alpha_{pl}$ is a power law index for the Planck function that depends on the temperature of the dust and the wavelengths of interest \citep[e.g,][]{holland03}.  Specifically,
\begin{equation}
\alpha_{pl} = \left| \frac{\rm log\Big(\frac{B_{\nu_{1}}}{B_{\nu_{2}}}\Big)}{\rm log\Big(\frac{\nu_{1}}{\nu_{2}}\Big)}\right|,
\end{equation}
where B$_{\nu}$ is the Planck Function and $\nu$ is the frequency. As a reference, a \cite{dohnanyi} collisional cascade will have $q\approx 3.5$, although other size distributions are possible depending on assumptions for the internal strength of the grains and other dynamical processes \citep[see][for a summary]{macgregor}.

Using this formalism and the flux from VLA 14A, we calculate $\alpha_{mm}=1.63$ \citep[consistent with ][]{macgregor} and a grain size distribution $q=2.83$, which would make HD 141569 a clear outlier from other discs with values $>3.0$.  As discussed in \cite{macgregor}, such a value for $q$  is inconsistent with proposed models for debris size distributions.  However, Fig.\,\ref{spec} highlights that deriving an appropriate $\alpha_{mm}$ is non-trivial for HD 141569.  

The ALMA 2.9 mm observations are not consistent with $\alpha_{mm}=1.63$ at $>2\sigma$ (where $\sigma$ is the total measurement uncertainty).  Furthermore, with the VLA 16A data, we find a 9 mm flux that is $\sim60\%$ lower than the VLA 14A observations, meaning the 9 mm flux density between semesters is inconsistent at $\gtrsim 2\sigma$.  Because of this, we argue that the additional ALMA and VLA observations support a steeper $\alpha_{\rm mm}$ than that found by \cite{macgregor}.

If only the two ALMA measurements at 870 $\mu$m and 2.9 mm are considered, then we find $\alpha_{\rm mm} = 1.81\pm 0.20$ (Fig.~\ref{spec}, dotted line).  This would correspond to a $q = 2.95\pm 0.11$, which while still shallow, is marginally consistent with some models for the grain size distributions in collisional cascades \citep[e.g.,][]{pan}.

Without a clear detection of any disc emission in the VLA datasets it is difficult to make firm conclusions regarding the true spectral index of the warm disc at cm wavelengths. We show in Fig.\,\ref{spec} the spectral slope that would be consistent with a \cite{dohnanyi} collisional cascade.  The red region is shown instead of a single line to highlight the range bounded between the two ALMA observations.  This is not intended to suggest that the spectra index must lie within this region; rather, it is intended to illustrate the range of possibilities with the current data set.

\section{Origin of the 9 mm Emission} 
 
\subsection{HD 141569A}

\begin{figure}
\centering
\includegraphics[width=0.5\textwidth]{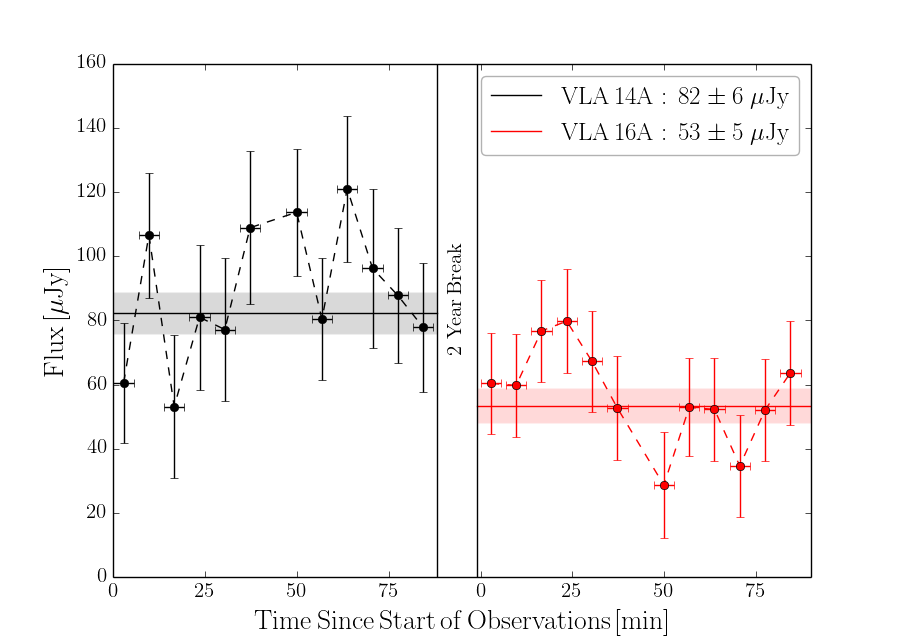}
\caption{Best fit 9 mm flux of HD 141569A as a function of on source time divided into $\sim5$ min chunks. \textbf{Left:} Semester 14A observations from June 2014 that found  $82 \pm 6~\rm \mu Jy$. \textbf{Right:} Semester 16A observations from June 2016 that found $53 \pm 5 ~\rm \mu Jy$. We use the {\scriptsize CASA} task \textit{uvmodelfit} and a point source model to fit the flux for each time chunk. Both observations achieved roughly 1 hour on source and have uncertainties given by the $\sigma_{\rm rms}$ of the images of each individual time chunk. The mean values of $\sigma_{\rm rms}$ are $21 ~ \mu \rm Jy~beam^{-1}$ and $16 ~ \mu \rm Jy~beam^{-1}$ for VLA 14A and VLA 16A, respectively. The solid lines represents the best fit flux for the entire length of each observation with the shaded region representing the total uncertainty.
\label{flux_time}}
\end{figure}

Fig.\,\ref{flux_time} shows a time series analysis of the HD 141569A 9 mm flux for both VLA 14A and VLA 16A. These points were generated by splitting out the data into $\sim5$ min chunks. Each time chunk was fit with a point source model using \textit{uvmodelfit} similar to the procedure in Section 3. The uncertainties are the $\sigma_{\rm rms}$ of the reconstructed image of each chunk. The mean values of $\sigma_{\rm rms}$ are $21 ~ \mu \rm Jy~beam^{-1}$ and $16 ~ \mu \rm Jy~beam^{-1}$ for VLA 14A and VLA 16A, respectively. While there is potential variability between the two years between observations, there is also potentially some low amplitude variability within each on $\sim1$ hr on source. However, we note  that there  is some correlation between the perceived variability in HD 141569A and HD 141569B, suggesting an additional systematic effect (Pearson correlation coefficient of $\sim0.69$). 

The VLA 16A observations presented here measure a flux density of  $53\pm 5~\mu$Jy  centred at the location of HD 141569A. If this flux is entirely due to stellar emission (i.e. no dust emission), and taking a stellar radius of$~\sim1.5$ R$_{\odot}$, then it would imply a brightness temperature of $\sim 5 \times 10^{6}$ K. This is nearly 500 times the photosphere temperature of HD 141569A (10500 K). While the photosphere temperature is not expected to be representative of the mm/cm brightness temperature of stars, the mm/cm temperature profile of A type stars is only just now being tested with observations \citep{white_sir}. The only star thoroughly studied and modelled at these wavelengths is the Sun, which at 1 cm has an observed brightness temperatures that is 2 - 3 times larger than the Solar photosphere temperature  \citep{loukitcheva}. Another example is TW Hydrae, a pre-MS K6 star that hosts an intricate circumstellar disc. VLA observations of TW Hydrae at 3.6 cm found more emission than expected from a simple extrapolation of the disk dust spectrum at shorter wavelengths \citep{wilner00}. Subsequent VLA observations showed no significant time variability and resolved this emission, ruling out a gyrosynchrotron origin in stellar activity and implicating a population of pebbles \citep{wilner05} and possibly also ionized wind \citep{pascucci}.

Thermal bremsstrahlung is a dominant emission mechanism in stars with $>~20\times10^{6}$ K \citep{gudel}. While the derived temperature is a factor of a few less than this, it could contribute to HD 141569A's emission. Magnetic fields can provide a strong, variable source of emission through synchrotron emission in stellar atmospheres, however, the magnetic field properties of Herbig Ae/Be stars (such as HD 141569A) are not very well constrained. Radio flaring has been observed in pre-MS stars of all spectral types and may be correlated to X-ray variability \citep{forbich}. Stellar winds are another possible source of radio emission, although the mass loss rate for A/B stars is thought to not be appreciable enough to contribute to significant radio emission \citep{drake}. If HD 141569A is in fact a $\lambda$ Boo type star \citep{murphy2015}, then accretion may be driving the observed radio variability. Ultimately, the stellar emission in the mm and cm from more massive stars, and particularly pre-MS stars, is poorly understood  \citep[see e.g.][]{gudel, cauley}.

\subsection{M Dwarfs}

\begin{figure}
\centering
\includegraphics[width=0.5\textwidth]{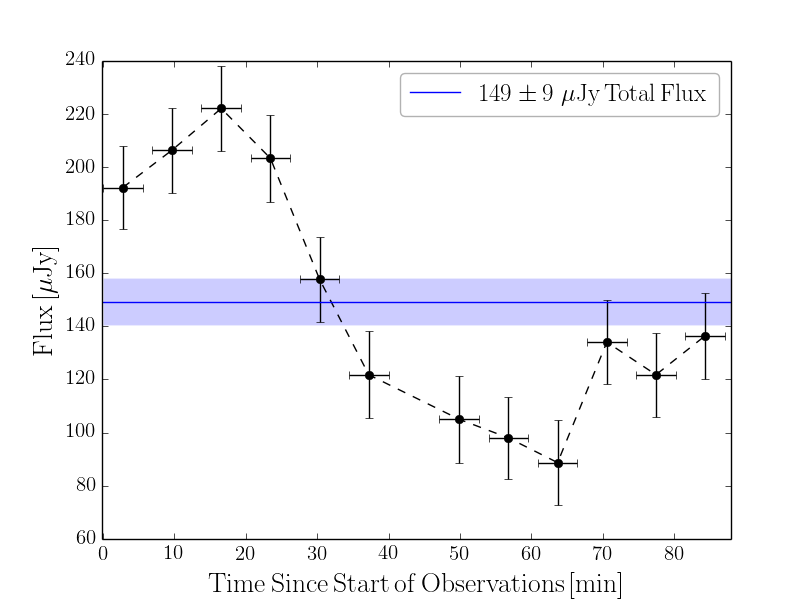}
\caption{Best fit 9 mm flux of HD 141569B from VLA 16A as a function of on source time divided into $\sim5$ min chunks. The observations found a total flux of $149 \pm 8 ~\rm \mu Jy$. The observations achieved roughly 1 hour on source and have uncertainties given by the $\sigma_{\rm rms}$ of the images of each individual time chunk. The mean value of $\sigma_{\rm rms}$ is $16 ~ \mu \rm Jy~beam^{-1}$. The solid line represents the best fit flux for the entire length of each observation with the shaded region representing the total uncertainty. HD 141569C was not detected and has an upper limit flux of $\sim5~\mu \rm Jy$.
}
\label{flux_timeB} 
\end{figure}

There are two candidate M dwarf companions to HD 141569A \citep{weinberger}. Their positions are denoted in Fig.\ref{fig1}. The peak of HD 141569B is at  the \textit{Gaia} predicted location and is consistent with a point source. There is a non-detection for the expected location of HD 141569C and the surrounding area. The effective temperatures of the stars are $3500\pm85$ K and $3200\pm85$ K for B and C, respectively \citep{weinberger}.

Fig.\,\ref{flux_timeB} shows a time series analysis of the HD 141569B 9 mm flux from VLA 16A. This plot was generated by the same procedure as Fig.\,\ref{flux_time}. The uncertainty of each data point is characterized by the $\sigma_{\rm rms}$ of reconstructed image of each time chunk. The mean $\sigma_{\rm rms}$ for all points is $16~\mu \rm Jy~beam^{-1}$. The flux density for  HD 141569B is measured to be $149\pm9~\mu$Jy (See Section 3). This is roughly 300\% of the $50~\mu$Jy flux recovered from VLA 14A \citep{macgregor}, although they were not able to clearly distinguish between B and C due to the large beam size.  The values from both semesters, nonetheless, imply very large brightness temperatures. If the two companion stars are indeed bound to HD 141569A, then they likely have the same approximate age of $\sim5$ Myr \citep{clampin}. These pre-MS M dwarfs would be expected to have variable radio emission \citep{forbich}, as is observed. The emission at these wavelengths may likely be dominated by magnetic effects, which will extend out to  $\sim3$ times R$_{*}$ with R$_{*} \approx \rm R_{Jupiter}$ \citep{burrows}. Using the VLA 16A flux density for HD 141569B and using the $\sigma_{\rm rms}$ as an upper level limit of HD 141569C, we find brightness temperatures of $6\times10^{8}$ K and $< 2\times10^{7}$ K, respectively. The derived temperatures are in line with the temperatures of other magnetically active M dwarfs \citep{burgasser}.

If the emission reported by \citet{macgregor} for the VLA 14A observations is equally distributed between HD 141569B and HD 141569C, then we expect a flux of $\sim 25~\mu$Jy for each M dwarf. This would correspond to a $>75\%$ drop in flux for HD 141569C and a $\sim500\%$ flux increase for HD 141569B. The flux of HD 141569B, however, is not constant throughout the 1.15 hr observations (see Fig.\,\ref{flux_timeB}) and ranges from about 90 to 220 $\mu$Jy. This variability of $>200\%$ is similar to what \citet{macgregor} observed in AU Mic. We note again however that there is some correlation between the perceived variability in HD 141569A and HD 141569B, likely due to systematic effects.  

The quiescent emission at these temperatures could be dominated by thermal bremsstrahlung \cite{gudel}. Possible flaring features have been observed in other M dwarfs with durations of less than 10 minutes \citep{burgasser}. Accurately characterizing the expected emission of these types of stars can also play an important role in debris disc studies. As HD 141569B and C were not detected in the either of the ALMA observations, this implies that they likely do not have significant, detectable debris and all the observed flux is stellar emission \citep{white}.

\section{Summary}

We have presented VLA Ka Band 9 mm (33 GHz) observations of the HD 141569 system. These 0.25 arcsec resolution observations targeted both the circumstellar disc around HD 141569A as well as the two M dwarf companions, HD 141569B and HD 141569C. The $4.7~\mu \rm Jy~beam^{-1}$ sensitivity was insufficient for detecting the disc.  We conclude that the previously constrained spectral index of $\alpha_{\rm mm} = 1.63$ is too shallow. Using ALMA 870 $\mu$m (345 GHz) and archival ALMA 2.9 mm (102 GHz) data, we place a lower level limit of $\alpha_{\rm mm} = 1.81\pm 0.20$, corresponding to a grain size distribution of $q = 2.95\pm 0.11$.

The recovered 9 mm emissions of $52\pm 5~\mu$Jy for HD 141569A and $149 \pm 9~\mu$Jy for HD 141569B are both consistent with point sources (HD 141569C was not detected). The brightness temperatures of HD 141569A and B are $\sim 5 \times 10^{6}$ K and $\sim 6\times10^{8}$ K, respectively. While there is clearly significant variability in the emission from HD 141569B, there is also non-negligible variability in HD 141569A.

\section*{Acknowledgments}

We thank the anonymous referee for comments that improved this manuscript.  J.A.W. and A.C.B. acknowledge support from an NSERC Discovery Grant, the Canadian Foundation for Innovation, The University of British Columbia, and the European Research Council (agreement number 320620). M.A.M. acknowledges support from a NSF Graduate Research Fellowship (DGE1144152) and under NSF Award No. 1701406. A.M.H. is supported by NSF grant AST-1412647.

The National Radio Astronomy Observatory is a facility of the National Science Foundation operated under cooperative agreement by Associated Universities, Inc. This paper makes use of the following ALMA data: ADS/JAO.ALMA[2013.1.00883.S]. ALMA is a partnership of ESO (representing its member states), NSF (USA) and NINS (Japan), together with NRC (Canada), NSC and ASIAA (Taiwan), and KASI (Republic of Korea), in cooperation with the Republic of Chile. The Joint ALMA Observatory is operated by ESO, AUI/NRAO and NAOJ.

%%%%%%%%%%%%%%%%%%%%%%%%%%%%%%%%%%%%%%%%%%%%%%%%%%

%%%%%%%%%%%%%%%%%%%% REFERENCES %%%%%%%%%%%%%%%%%%

% The best way to enter references is to use BibTeX:

%\bibliographystyle{mnras}
%\bibliography{example} % if your bibtex file is called example.bib

% Alternatively you could enter them by hand, like this:
% This method is tedious and prone to error if you have lots of references

%%%%%%%%%%%%%%%%%%%%%%%%%%%%%%%%%%%%%%%%%%%%%%%%%%

%%%%%%%%%%%%%%%%% APPENDICES %%%%%%%%%%%%%%%%%%%%%

%\appendix
%
%\section{Some extra material}
%
%If you want to present additional material which would interrupt the flow of the main paper,
%it can be placed in an Appendix which appears after the list of references.

%%%%%%%%%%%%%%%%%%%%%%%%%%%%%%%%%%%%%%%%%%%%%%%%%%

% Don't change these lines
\bsp	% typesetting comment
\label{lastpage}
\end{document}